\documentclass[aps,pra,twocolumn,showpacs,superscriptaddress]{revtex4}

\usepackage{graphicx}
\usepackage{amsmath}

\begin{document}
\title{Test-State Approach to the Quantum Search Problem}

\author{Arun Sehrawat}
\email[email: ]{arun02@nus.edu.sg}
\affiliation{Centre for Quantum Technologies, %
National University of Singapore, Singapore 117543, Singapore}

\author{Le Huy Nguyen}
\affiliation{Centre for Quantum Technologies, %
National University of Singapore, Singapore 117543, Singapore}
\affiliation{Graduate School for Integrative Sciences and Engineering, %
National University of Singapore, Singapore 117597, Singapore}

\author{Berthold-Georg Englert}
\affiliation{Centre for Quantum Technologies, %
National University of Singapore, Singapore 117543, Singapore}
\affiliation{Department of Physics, %
National University of Singapore, Singapore 117542, Singapore}

\date{17 February 2011}

\begin{abstract}
The search for ``a quantum needle in a quantum haystack'' is a 
metaphor for the problem of finding out which one of a permissible set of
unitary mappings---the oracles---is implemented by a given black box.
Grover's algorithm solves this problem with quadratic speed-up as compared
with the analogous search for ``a classical needle in a classical haystack.''
Since the outcome of Grover's algorithm is probabilistic---it gives the
correct answer with high probability, not with certainty---the answer requires
verification.
For this purpose we introduce specific test states, one for each oracle.
These test states can also be used to realize ``a classical search for the
quantum needle'' which is deterministic---it always gives a definite answer
after a finite number of steps---and faster by a factor of $3.41$ than the
purely classical search.
Since the test-state search and Grover's algorithm look for the same quantum
needle, the average number of oracle queries of the test-state search is the
classical benchmark for Grover's algorithm. 
\end{abstract}
\pacs{03.67.Ac}

\maketitle

\section{Introduction}\label{sec:Intro}
In recent decades, the quantum theory found another practical use in the field
of quantum information processing: 
Quantum computation \cite{QC1, QC2, QC3} and quantum information theory
\cite{QI1, QI2} are extensively researched by several scientific communities. 
On one hand, the superposition principle of quantum mechanics speeds up the
computation for important classes of computational problems. 
On the other hand, entanglement assists us in sending information from one
place to another in a secure way. 
Both branches---quantum computation and quantum information theory---are
interlinked and are growing rapidly.  

``Searching an item in a given database'' is a well-known computational
problem which has been studied with different conditions in both classical
\cite{CC} and quantum \cite{GA} contexts. 
When an unsorted database stored in the memory of a classical computer (CC) is
given, then the average number of iterations required by the CC to complete
this \emph{classical search} grows linearly with the total number of items
present in the database. 
But, if the items are already present in an order in the memory, then the
average number of iterations scales up logarithmically with respect to the
total number of items.  

Grover introduced a \emph{quantum search algorithm} \cite{GA} for an analogous
``quantum search problem:'' A quantum computer (QC) searches a particular ket
out of a set of kets from the computational basis.
Or, more precisely, the quantum search finds which one of a set of unitary
mappings---the \emph{oracles}---is implemented by a black box. 
One can tackle this problem like its classical analog by testing
for each oracle one by one in a sequence till the match is found. 
Alternatively, we can exploit the superposition principle and address all the
oracles simultaneously, with Grover's search algorithm (GA).  
This gives a quadratic speedup to GA in comparison with the classical search.

But the answer returned by GA is the correct one with a high probability only,
not with certainty.
It is, therefore, necessary to verify the answer.
This verification is done with the aid of the \emph{test states} that we
introduce here, one test state for each oracle.

The test states can also be used for a classical-type search of the quantum
data set (that is, the set of oracles). 
Such a test-state search is deterministic---it will give the correct answer
after a finite number of queries---and the average number of queries is
proportional to the number of permissible oracles, but fewer by a factor of
$3.41$ than the average number of queries for the corresponding classical
search of an unstructured classical database.

A single iteration of the test-state approach is a three-step process. 
First, we prepare a test state, which is a certain superposition of all the
kets of the set under search. 
Second, we pass it through the oracle---the very same oracle that is employed
by GA.  
Finally, we retrieve the information by a measurement on the processed test 
state. 
As is the case in the classical search, this measurement says ``yes'' or
``no'' if the test state matches the oracle or not.
In marked contrast to the classical search problem, however,
there are different ``no'' answers depending on the actual oracle, 
and the measurement extract the available information.
The choice of test state for the next round is then guided by this
information, and this guidance leads to a substantial reduction of the average
number of trials needed before successful termination of the search.

The structure of this article is as follows. 
Section~\ref{sec:SP} comprises of the definitions and the algorithms for
the search problem in both classical and quantum contexts. 
In Sec.~\ref{sec:TSA}, a comprehensive description of the test-state approach
to the quantum search problem is provided. 
GA with test-state verification is then discussed in Sec.~\ref{sec:GA-TSV},
and Sec.~\ref{sec:TSS-alt} deals with alternative test-state search strategies.
In Sec.~\ref{sec:EXPT}, we describe the quantum circuit for the construction
of the test state and for realizing the measurements. 
We conclude with a summary and discussion in Sec.~\ref{sec:SummDiss}, 
and two appendixes contain additional material.

\section{Search problem}\label{sec:SP} 
Suppose someone gives us a list of one hundred names of different animals on a
piece of paper, and ask where ``Lion'' appears on this list. 
If ``Lion'' appears exactly once on the list, and the list is not ordered
in any obvious way, then we have to go through about fifty names on average 
before we find ``Lion.'' 
For a search of this kind, neither a CC nor a QC can directly helps us, because
the data (names) are given on a piece of paper.  

In order to use a CC or a QC for this kind of database search, first we have
to convert the data into an accessible format. 
For example, in case of a CC for such a search, first we have to
load the data (the given list) into the memory of a CC. 
However, we can find the name Lion in the process of converting the list of
names into an electronic format (in terms of strings of bits) and storing them
in the memory. 
So, neither a CC nor a QC is very helpful for a search of this kind. 
In other words, a CC (QC) is helpful for a database search only when the
database is given in an electronic format (quantum format). 

Furthermore, a QC also cannot search a classical database without a ``quantum
addressing scheme'' \cite{QCQI} where the classical database is converted into
a quantum format (in terms of quantum kets). 
So, the process of searching a marked string of bits with a CC in a classical
database which is stored in the memory of a computer is called as 
\emph{classical search}. 
Similarly, \emph{quantum search} is a process where a QC searches a marked
quantum state (or, rather, a particular unitary operation) out of a set of
quantum states 
(or, rather, a set of unitary operations). 
Classical and quantum searches are analogous but not the same, their detailed
description is given in the following.

\subsection{Classical search}\label{sec:CA}
Suppose we have an unsorted database as a set
\begin{equation}
\mathcal S^{N}_{C}\equiv\{0,\cdots,j,\cdots,N-1\}
\label{set}
\end{equation}
of a total of $N$ items stored electronically in the memory of a CC. 
Each item is labeled by an index from 0 to $N-1$ and further represented  by
a $n$-bit string in binary representation. 
For convenience, we shall confine our attention to the case 
$N = 2^{n}$ ($0 \equiv 00\cdots0$, $N-1 \equiv 11\cdots1$), 
but the following algorithms and the test-state approach can be implemented
for an arbitrary value of $N$.  

Throughout the article we shall consider only the case of a single matching
item.
The task of the search problem is to recover the corresponding index ($n$-bit
string) to the marked item at the end of computation. 

The method employed by a CC to solve the search problem is by checking every
element of $\mathcal S^{N}_{C}$ one by one in a sequence till a match is found
\cite{CC}. 
A single iteration of this classical algorithm is a three-step process given
as follows. 
\textbf{Step~1:} The CC picks a $n$-bit string at random from the set
$\mathcal S^{N}_{C}$ as an input. 
\textbf{Step~2:} The CC checks whether or not this string matches with our
query. 
\textbf{Step~3:} It produces an answer to the question in terms of ``yes''
or ``no.'' 
If the answer is ``yes,'' then the CC stops the computation and produces the
string as the result, and the corresponding item will be the matching item. 
If the answer turns out to be ``no,'' then the CC picks another string at 
random from the set $\mathcal S^{N}_{C}$ as an input, with items tested earlier
excluded, and asks the same question. 
If the answer is again ``no,'' then the CC repeats the above procedure until
it hits the matching item. 
One of the main points in this classical algorithm is, 
``Every time the CC picks at random only one $n$-bit string, and its current
guess does not depend on previous guesses'' other than excluding them.
In this way, a CC needs, on average, as many as 
\begin{equation}
\mathit{G}_{C}(N)=\frac{N+1}{2}-\frac{1}{N}\label{Gc}
\end{equation}
queries of the database before it finds the matching item.
This is an immediate consequence of the recurrence relation
\begin{equation}
  \label{GC-recurr}
 \mathit{G}_C(N+1)=1+\frac{N}{N+1}G_C(N)\quad\mbox{for}\enskip N>1
\end{equation}
that commences with $G_C(1)=0$.

Since $\mathit{G}_{C}(N)\propto N$ for $N\gg1$, this classical search algorithm
is linear in the number of candidate items.
If, rather than being unstructured, the data were sorted beforehand, then the
problem could be completed by a binary search in approximately $\log_{2}N$
iterations \cite{CC}.

\subsection{Quantum search}\label{sec:GA} 
In this section, an analogous quantum search problem to the classical one and
a brief description of GA \cite{GA} is provided. 
Throughout the text we represent the single-qubit Pauli vector operator 
$\vec{\sigma}$ by $\left(X, Y, Z \right)$ and the identity operator by $I$.  

In the step from classical to quantum, bits are replaced by qubits. 
So, for each index ($n$-bit string) $j$ of $\mathcal S^{N}_{C}$ defined by
Eq.~(\ref{set}) there exist a $n$-qubit quantum ket $|j\rangle$, the so-called
\emph{index ket}. 
There is then a unitary operation $\mathcal{O}^{j}$---the $j$th
\emph{oracle}---which gives a conditional phase shift of $\pi$ to the index
ket $|j\rangle$ only,   
\begin{equation}\label{Oj}
\mathcal{O}^{j}\equiv{(-\mathcal{I})}^{|j \rangle \langle j|}
=\mathcal{I}-2|j \rangle \langle j|\,,
\end{equation}
where $\mathcal{I}=I^{\otimes n}$ is the identity operator in the
$N$-dimensional Hilbert space.   

One can define an analogous quantum search problem to the classical search
problem of Sec.~\ref{sec:CA} in the following way: 
Suppose someone gives us a quantum black box, which is implementing one of the
$N$ different oracles, and asks us to find out which of the oracles is the
case without actually opening the box and looking inside. 
Clearly, we are not using QC to search a marked item in a classical database,
but we are searching the index ket corresponding to the given oracle.
The question of how many queries of the database are now needed, reads 
``How many times must one use the quantum black box to find out the correct
result?''   

The most efficient way of finding out which oracle is the actual one is
GA \cite{optGA}.
GA begins by applying the Hadamard gate
\begin{equation}
H=(X+Z)/\sqrt{2} \label{H}
\end{equation}
to each qubit, after initially preparing the state with index ket
$|0\rangle \equiv |0\rangle^{\otimes n}$. 
The operation $H^{\otimes n}$ creates a superposition of all the index kets of
$\mathcal S^{N}_{Q}$ with equal amplitude $1/\sqrt{N}$. 
The next step is an application of the Grover iteration operator
$\mathcal{G}$, geometrically it is a
rotation composed of two reflection operations as
$\mathcal{G}=\mathcal{D}\mathcal{O}$. 
The operator $\mathcal{O}$ is the same quantum oracle (black box) defined by
Eq.~(\ref{Oj}), whose unknown index we have to find. 
The diffusion operator $\mathcal{D}$ gives an inversion about the average
\cite{GA},  
\begin{equation}
\mathcal{D}=-H^{\otimes n}\bigl(\mathcal{I}-2 |0\rangle\langle 0|\bigr)
             H^{\otimes n};\label{D}
\end{equation} 
its central piece is the $0$th oracle $\mathcal{O}^0$.
 
GA is probabilistic in nature in the sense that, after applying $\mathcal{G}$
several times, the probability of the privileged index ket becomes significantly
higher than the probabilities of the other index kets. 
Finally, we read out the output by performing projective measurements on each
qubit, and so find one of the index kets.
After $k$ applications of  $\mathcal{G}$, we have \cite{GA}
\begin{equation}
  \label{GAprobNk}
  p^{(N)}_k=\sin\bigl((2k+1)\theta_N\bigr)^2\quad\mbox{with}\enskip
  \sin\theta_N=1/\sqrt{N}
\end{equation}
for the probability that the oracle associated with the final output state 
is the one which the black box is executing. 
Upon optimizing $k$, 
GA solves the quantum search problem by using the black box only 
\begin{equation}
\mathit{G}_{Q}(N)= 0.69\sqrt{N} \label{Gq}
\end{equation}
times when ${N\gg1}$; see Sec.~\ref{sec:GA-TSV} below. 
The quadratic speedup of $\mathit{G}_Q(N)\propto\sqrt{N}$ versus  
$\mathit{G}_C(N)\propto{N}$ is owed to the computational power of quantum
physics; specifically, the superposition principle is at work.
We emphasize that the outcome of GA is not guaranteed to be the correct
answer; it can be incorrect with a probability that is very small but
definitely nonzero. 

In passing, we note the following.
A general treatment of GA for multiple targets and for an arbitrary value of
$N$ is given in Ref.~\cite{multGA}. 
Moreover, GA is a special case of the quantum amplitude
amplification~\cite{QAA}. 
In addition, one can get rid of the probabilistic nature of GA if one has the
option of changing the structure of the diffusion operator $\mathcal{D}$ and
the oracle $\mathcal{O}$ \cite{extGA}. 
When one is only allowed to use the given black box, namely the oracles of
Eq.~(\ref{Oj}), but not to look inside and change the setting, then GA remains
probabilistic in nature. 

So, one needs a confirmation step to be sure of the result obtained by GA. 
A single iteration of the test-state search introduced in the next
section acts as a confirmation step for GA, where the verification matter is
discussed after Eqs.~(\ref{ab}).  
Details of GA with test-state verification are given in Sec.~\ref{sec:GA-TSV}.

\section{Test-state search}\label{sec:TSA} 
In this section, we introduce the test-state approach to the quantum search
problem described in Sec.~\ref{sec:GA}---where one has to identify the actual
oracle which is implemented by the given black box. 
The features of both classical and quantum approaches are embodied in this
approach. 
A single iteration in the test-state approach can be summarized in
the following three steps. 

\textbf{Step~1:} We pick an index ket $|j\rangle$ and
prepare the corresponding test state. 
\textbf{Step~2:} We pass the test state through the given quantum black box
which is executing one of the oracles of Eq.~(\ref{Oj}). 
\textbf{Step~3:} We extract the information with the help of a particular
probability-operator measurement (POM) \cite{POVM1, POVM2}. 
Here, a ``single iteration'' comprises of these three steps, which are similar
to the classical search algorithm of Sec.~\ref{sec:CA}. 
The result of the POM gives an answer to the same question---whether or not
the black box is executing the oracle $\mathcal{O}^{j}$---in terms of ``yes''
or ``no.'' 
The answer ``yes'' tells us that the black box is executing the
corresponding oracle to the index ket we picked, and we terminate the search.

Even if the answer is ``no,'' the result of the POM gives us some information
about the actual oracle. 
This information facilitates an educated guess and a judicious choice of the
test state for the next iteration.

The correct result is obtained after a finite number of iterations.
In other words, the test-state search is deterministic, rather than
probabilistic.
And, the systematic educated guessing makes the test-state search more
efficient than a truly classical search, in which all test states would be
chosen at random: 
For $N\gg1$, the test-state search needs fewer guesses by a factor of $3.41$.

\subsection{A single iteration in the test-state search}\label{sec:itrTSA} 
In this section, we construct the test states for verification of the outcome
of GA and discuss the three steps of one iteration round in the test-state
approach to determining the actual oracle of the quantum search problem.
The narrative follows the steps in sequence.  

\textbf{Step~1---Preparing the test state:}
We pick an index ket $|j\rangle$ from the set
\begin{equation}
  \label{setQ}
  S^{N}_{Q}=\bigl\{|0\rangle,\dots,|j\rangle,\dots,|N-1\rangle\bigr\}
\end{equation}
of all index kets.
For the very first round of iteration, the choice of $|j\rangle$ is random,
but for all subsequent rounds the choice is dictated by the result of the
measurement in Step~3, as discussed in Sec.~\ref{sec:CP}.

Then we prepare the corresponding test state $|t_j\rangle$ which is of the form
\begin{equation}  \label{tj}
|t_j\rangle = a|j\rangle+b\sum_{l(\neq j)}|l\rangle\,,
\end{equation} 
where $a$ is the amplitude of the privileged index ket $|j\rangle$ and $b$ is
the common amplitude of all other index kets. 
Both $a$ and $b$ are functions of $N$; it suffices to consider only real
positive values for $a$ and $b$, but this is a restriction of convenience, not
of necessity. 

In Sec.~\ref{sec:QCR}, we present a quantum circuit for constructing the test
state $|t_0\rangle$. 
The test state $|t_0\rangle$ can be transformed into any other test state
$|t_j\rangle$ by applying the $X$ operations on the relevant qubits. In other
words, each $|t_j\rangle$ is equivalent to $|t_0\rangle$ up to some
single-qubit operations.  

\textbf{Step~2---Processing the test state:}
We pass the test state $|t_j\rangle$ through the given quantum black box. 
We recall that the black box is implementing one of the $N$ different oracles
of Eq.~(\ref{Oj}), but we do not know which oracle is the case. 
If the black box is implementing the $j$th oracle, then the resultant state
is   
\begin{equation} \mathcal{O}^j|t_j\rangle=
|t_j^j\rangle =-a|j\rangle+b\sum_{l(\neq j)}|l\rangle. \label{tjj} 
\end{equation}
If the black box is not implementing the $j$th oracle, but some other one, the
$k$th oracle, say, then the resultant state is
\begin{equation} 
k\neq j\,:\quad
\mathcal{O}^{k}|t_{j}\rangle=|t_{j}^{k}\rangle 
=a|j\rangle-b|k\rangle+b\sum_{l(\neq j,k)}|l\rangle. \label{tjk}  
\end{equation}
Result $|t_j^j\rangle$ says ``yes, it is the $j$th oracle'' whereas each
$|t_{j}^{k}\rangle$ with $k\neq j$ says ``no, it is not the $j$th oracle,'' and
we note that there is one ``yes'' but $N-1$ different ``no''s.

We define the ``no'' set $\mathcal{C}_{j}^{N}$ to index ket $|j\rangle$ as the
collection of all $N-1$ ``no'' states of Eq.~(\ref{tjk}),
\begin{equation}
\mathcal{C}_{j}^{N}=\bigl\{|t_{j}^{0}\rangle,\cdots,|t_{j}^{j-1}\rangle,
|t_{j}^{j+1}\rangle, \cdots,|t_{j}^{N-1}\rangle\bigr\}.\label{C}
\end{equation}
In order to be able to distinguish the ``yes'' ket $|t_{j}^{j}\rangle$ from
the ``no'' kets in $\mathcal{C}_{j}^{N}$, we demand that
\begin{equation} 
\langle t_{j}^{k}\arrowvert t_{j}^{j}\rangle =0\quad\mbox{for $k\neq j$},
\label{tt} 
\end{equation}
so that the ``yes'' ket is orthogonal to all ``no'' kets.
Together with the normalization of the test-state ket $|t_j^j\rangle$, this
gives 
\begin{eqnarray}
a & = & \sqrt{(N-3)/(2N-4)}, \nonumber \\
b & = & \sqrt{1/(2N-4)}, \label{ab}
\end{eqnarray}
for the amplitudes in Eq.~(\ref{tj}).

The use of the test states for the verification of the outcome of GA, is quite
obvious: 
After GA identifies the $j$th oracle, we prepare the $j$th test state
$|t_j\rangle$ and let the oracle act on it. 
Then we perform a measurement that determines whether the resulting ket is
proportional to the ``yes'' ket $|t_j^j\rangle$ or resides in the orthogonal
subspace spanned by the ${N-1}$ ``no'' kets. 
If we find the ``yes'' ket, the search is over; otherwise, we have to execute
GA another time. 
An alternative confirmation step for GA, where one has to use the black box at
most two times, is described in Appendix~\ref{sec:ACS}. 

As Eqs.~(\ref{ab}) show, there are test states for $N>2$, but none for $N=2$.
This is as it should be.
For, the two $N=2$ oracles $\mathcal{O}^0=|1\rangle\langle1|-|0\rangle\langle0|$
and $\mathcal{O}^1=|0\rangle\langle0|-|1\rangle\langle1|$ are simply
indistinguishable; they do not tell the index kets $|0\rangle$ and $|1\rangle$
apart. 

Turning our attention to the ``no'' kets, we observe that they are the edges
of a $(N-1)$-dimensional pyramid,
\begin{equation}
k\neq j,\ l\neq j\,:\quad\langle t_{j}^{k}\arrowvert t_{j}^{l}\rangle
=\lambda + (1-\lambda)\delta_{kl} \label{kl}
\end{equation}
with $\lambda =(N-4)/(N-2)$.
In the terminology of Ref.~\cite{pyramid3}, the pyramid is acute ($\lambda>0$) 
for $N>4$, orthogonal ($\lambda=0$) for $N=4$, and flat ($\lambda=-1/(N-2)$) 
for $N=3$.

The case $N=4$ is particular: 
We have $a=b=1/2$ and all four test states are identical.
The ``no'' states for one index ket are pairwise orthogonal; 
they are ``yes'' states for the other index kets.
As a consequence, testing the oracle once with the one common test state will
reveal its identity. 

This observation is sometimes stated as ``GA needs to query the oracle only
once for $N=4$.'' 
Indeed, we have $p^{(4)}_1=1$ in Eq.~(\ref{GAprobNk}).
This peculiarity of GA comes about because the common ${N=4}$ test state is
also the ${N=4}$ initial state of GA, and the ${N=4}$ version of the diffusion
operator $\mathcal{D}$ of Eq.~(\ref{D}) maps the SRM kets of Eq.~(\ref{Tjk3})
below onto the computational basis, in which the outcome of GA is obtained. 
 
\textbf{Step~3---Measuring the result:}
When measuring the state that results from applying the black-box oracle to
the $j$th test state $|t_j\rangle$, we not only need to distinguish between
``yes'' and ``no'' but also want to acquire information about which of the
``no''s is the case, so that we can make a judicious choice for the next test
state. 
Thanks to the pyramidal structure of the ``no'' kets, the POM that maximizes
our odds of guessing right is the so-called square-root measurement (SRM)
\cite{SRM,pyramid3}.

For $N=3$, there is no useful POM of this kind because the two ``no'' states
are the same, as is exemplified by $|t_0^1\rangle=-|t_0^2\rangle$.
For $N>3$, the SRM
\begin{equation}
  \sum_{k=0}^{N-1}\Pi_{j}^{k}=\mathcal{I} \label{DECOM}
\end{equation}
has the rank-1 outcomes $\Pi_{j}^{k}=| T_{j}^{k}\rangle\langle T_{j}^{k}|$
with
\begin{eqnarray}\label{Tjk3}
 | T_{j}^{k}\rangle  &=&
{\left(\sum_{l=0}^{N-1}| t_{j}^{l}\rangle\langle t_{j}^{l}|\right)}^{-1/2}
| t_{j}^{k}\rangle\nonumber\\
&=&\left\{
  \begin{array}{l}\rule{0pt}{3ex}
    | t_{j}^{j}\rangle\quad\mbox{for}\enskip k=j\\[1ex]\displaystyle
    b|j\rangle-x|k\rangle+y\sum_{l(\neq j,k)}|l\rangle
    \quad\mbox{for}\enskip k\neq j  
  \end{array}\right.
\end{eqnarray}
where
\begin{equation}
  \label{xy}
  y=\frac{1+a}{N-1},\quad x=1-y
\end{equation}
and $a,b$ are the coefficients of Eqs.~(\ref{tj})--(\ref{tjk}) and (\ref{ab}).

Since
\begin{equation}
  \label{TjkTjl}
  \langle T_{j}^{k} | T_{j}^{l}\rangle =\delta_{kl},
\end{equation}
the SRM is an orthogonal measurement, a standard von Neumann measurement, 
not a POM proper.
Therefore, the SRM can be implemented by a unitary transformation followed by
measuring the computational basis.
One quantum circuit for such a unitary transformation is given in
Sec.~\ref{sec:SRM}.

\subsection{Conditional probabilities}\label{sec:CP}
The probability of getting the $l$th outcome if the processed $j$th test state
is $|t_j^k\rangle$ is given by 
\begin{equation} 
\mathrm{prob}(t_{j}^{k}\to\Pi_{j}^{l})
=\langle t_{j}^{k} | \Pi_{j}^{l} | t_{j}^{k}\rangle 
=\left|\langle T_{j}^{l}|t_{j}^{k}\rangle\right|^{2}.
\end{equation}
It follows from Eqs.~(\ref{tjj}), (\ref{tjk}), and (\ref{Tjk3}) that there are
three cases,
\begin{equation} 
\mathrm{prob}(t_{j}^{k}\to\Pi_{j}^{l})=
\left\{
\begin{array}{l@{\ \mbox{for}\ }r}
1&              k=j,\ l=k\\
\alpha_{N-1}&   k\neq j,\ l=k\\
\beta_{N-1}&    k\neq j,\ l\neq k
\end{array}
\right.\label{p}
\end{equation}
where
\begin{eqnarray}\label{alpha-beta}
\beta_{N-1}&=&\frac{1}{(N-1)^{2}}
{\left(\sqrt{N-3}-\frac{\sqrt 2}{\sqrt{N-2}}\right)}^{2},
\nonumber\\
\alpha_{N-1}&=& 1-(N-2)\beta_{N-1}\nonumber\\
&=& \frac{1}{(N-1)^{2}}{\left(\sqrt{N-3}+\sqrt{2N-4}\right)}^{2},
\end{eqnarray}
with the subscript ${N-1}$ stating the number of different ``no'' outcomes.

The first case in Eq.~(\ref{p}) is the affirmative ``yes, it is the $j$th
oracle'' answer that terminates the search.
The second and third cases both say ``no, it is not the $j$th oracle.''
Thereby, the probability $\alpha_{N-1}$ of getting the $k$th outcome when the
black box implements the $k$th oracle is larger than the probability
$\beta_{N-1}$ for all other ``no'' outcomes.
Upon finding the $l$th outcome, we will therefore guess that the black box
contains the $l$th oracle and choose $|t_l\rangle$ as the next test state.
The choice of SRM maximizes the probability that this educated guess is right.

After the first wrong guess $|j\rangle$, we exclude the index ket $|j\rangle$ 
from the list of candidates, and have the set 
\begin{equation}
  \mathcal S^{N-1}_{Q}=\bigl\{|0\rangle,\cdots,|j-1\rangle,|j+1\rangle,
                             \cdots,|N-1\rangle\bigr\}
\end{equation}
of the remaining $N-1$ index kets for the next round.
Having found SRM outcome $\Pi_j^l$, we repeat the iteration described in
Sec.~\ref{sec:itrTSA} on the set $\mathcal S^{N-1}_{Q}$ by taking the index
ket $|l\rangle$ as the next educated guess, for which the ``no'' probabilities
are $\alpha_{N-2}$ and $\beta_{N-2}$. 
If this guess is also wrong, then the $l$th index ket can be excluded as well,
and we are left with $N-2$ candidates and a new educated guess for the next
test state with ``no'' probabilities $\alpha_{N-3}$ and $\beta_{N-3}$.
And so forth, until we either get the ``yes'' answer, or we are left with four
candidates only, having excluded $N-4$ index kets successively.
The common test state for $N=4$ will then surely give us the ``yes'' answer;
in the present context, this is confirmed by $\alpha_3=1$ and $\beta_3=0$ in
Eqs.~(\ref{p}) and (\ref{alpha-beta}).  

In each round of iteration in the test-state search, we are using the black
box once. 
Accordingly, the average number of oracle queries before a ``yes'' answer is
obtained, is given by 
\begin{equation}\label{Gt}
  \mathit{G}_T(N)=p^{(N)}_1+2p^{(N)}_2+3p^{(N)}_3+\dots+(N-3)p^{(N)}_{N-3}
\end{equation}
where $p^{(N)}_m$ is the probability that the search terminates after the $m$th
round.
For $N>4$, these probabilities are
\begin{eqnarray}\label{pm}
p^{(N)}_1& = & \frac{1}{N},\nonumber\\
p^{(N)}_2&=&(1-p^{(N)}_1)\alpha_{N-1}=\frac{N-1}{N}\alpha_{N-1},\nonumber\\
p^{(N)}_3&=&(1-p^{(N)}_1-p^{(N)}_2)\alpha_{N-2}\nonumber\\
        &=&\frac{N-1}{N}(1-\alpha_{N-1})\alpha_{N-2},\nonumber\\
&\vdots&\\
p^{(N)}_{N-4}&=&(1-p^{(N)}_1-p^{(N)}_2-\cdots-p^{(N)}_{N-5})\alpha_5\nonumber\\
           &=&\frac{N-1}{N}(1-\alpha_{N-1})\cdots(1-\alpha_{6})\alpha_{5}
\nonumber\\
p^{(N)}_{N-3}&=&1-p^{(N)}_1-p^{(N)}_2-\cdots-p^{(N)}_{N-4}\nonumber\\  \nonumber
           &=&\frac{N-1}{N}(1-\alpha_{N-1})\cdots(1-\alpha_{6})(1-\alpha_{5}).
\end{eqnarray}
Without the educated guesses provided by the SRM, one would have to resort to
choosing the test state for the next iteration at random, just as one does in
a purely classical search, which amounts to the replacement $\alpha_L\to1/L$
and yields $p^{(N)}_1=p^{(N)}_2=\cdots=p^{(N)}_{N-4}=1/N$, $p^{(N)}_{N-3}=4/N$.
But with the systematic educated guesses, we have
\begin{equation}
  \label{alphan}
  \alpha_L\approx\frac{3+\sqrt{8}}{L}\quad\mbox{for}\enskip L\gg1\,,
\end{equation}
and the probabilities for early termination are substantially larger than $1/N$.

Equations (\ref{Gt}) and (\ref{pm}) yield the recurrence relation
\begin{equation}
  \label{GT-recurr}
  G_T(N+1)=1+\frac{N}{N+1}(\alpha_N-\beta_N)+\frac{N^2\beta_N}{N+1}G_T(N),
\end{equation}
which commences with $G_T(4)=1$ and reduces, as it should, 
to its $G_C$ analog in Eq.~(\ref{GC-recurr}) for $\alpha_N=\beta_N=1/N$.
With the aid of the large-$L$ form of $\alpha_L$ in Eq.~(\ref{alphan}) and the
corresponding statement for $\beta_L$, we then find that the average number of
queries in the test-state search is given by
\begin{equation}
  \label{eq:GTapprox}
  \mathit{G}_T(N)\approx\frac{N}{4+\sqrt{8}}=\frac{N}{6.83}
\quad\mbox{for}\enskip N\gg1\,.
\end{equation}
The comparison with the classical search,
\begin{equation}
  \frac{\mathit{G}_T(N)}{\mathit{G}_C(N)}\approx\frac{1}{2+\sqrt{2}}
     =\frac{1}{3.41}\quad\mbox{for}\enskip N\gg1,
\end{equation}
shows that the judicious choice of the next test state has a substantial
pay-off: We need much fewer queries.

\begin{figure}
\centerline{%
\includegraphics[viewport=61 573 291 736,clip=true]{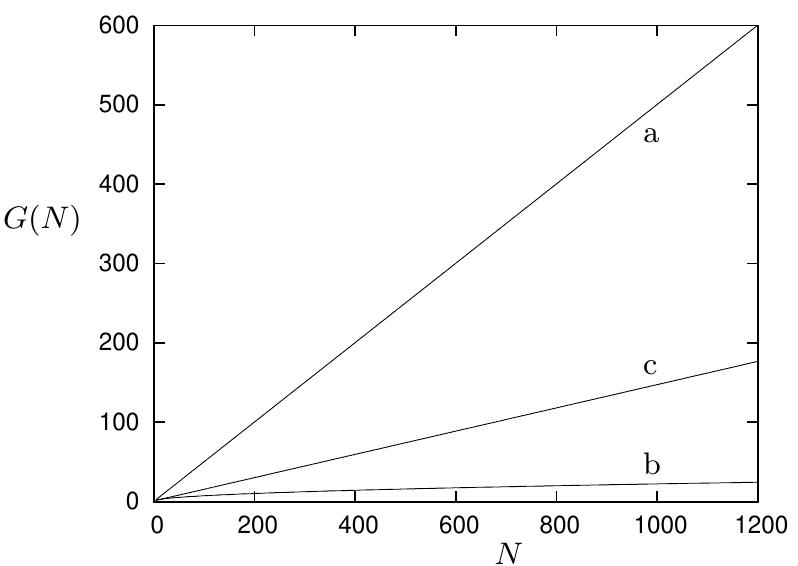}
}
\caption{\label{fig:1} 
  Average number $G(N)$ of oracle queries as a function of  the
  total number $N$ of index kets.
  Curve ``a'' shows $G_C(N)$ of Eq.~(\ref{Gc}) for the classical search
  strategy.
  Curve ``b'' shows $G_Q(N)$ of Eq.~(\ref{GQ-opt}) for Grover's search
  algorithm, supplemented by test-state verification and optimized for least
  number of queries per search cycle. 
  Curve ``c'' shows $G_T(N)$ of Eq.~(\ref{GT-recurr}) for the test-state
  search.   
}
\end{figure}

Since the test-state search and GA are both determining the
actual oracle inside the quantum black box, the classical-type ``yes/no''
approach of the test-state search sets the benchmark for the quantum search
with GA. 
It is true, that both $G_C(N)$ and $G_T(N)$ grow linearly with the number $N$ of
candidate items, whereas $G_Q(N)$ grows proportional to $\sqrt{N}$---and this
quadratic speed-up is, of course, the striking advantage of the quantum search
algorithm---but the reduction of the average number of queries by the factor of
$3.41$ is truly remarkable by itself.
It, too, is a benefit of the superposition principle. 
The three search strategies are compared in Fig.~\ref{fig:1}, which shows 
$G_C(N)$, $G_T(N)$, and $G_Q(N)$ as functions of $N$.

\section{Grover's algorithm with test-state verification}\label{sec:GA-TSV}
As recalled in Sec.~\ref{sec:GA} above, a single GA cycle consists of the
preparation of the initial state, $k$ applications of
${\mathcal{G}=\mathcal{DO}}$, followed by a measurement in the computational
basis that is composed of the index kets.
After the measurement finds index ket $|j\rangle$, we apply the oracle to
test state $|t_j\rangle$, measure the resulting state with the SRM,
and so decide whether the actual oracle is $\mathcal{O}^j$ or not. 
The search terminates when this test says ``yes.''
But if the reply is ``no,'' we execute another GA cycle.

The probability that a GA cycle finds the correct index state is $p_k^{(N)}$
of Eq.~(\ref{GAprobNk}).
It follows that the probability that the search terminates after the $m$th
cycle is
\begin{eqnarray}
  \label{probGAterm}
  \bigl(1-p_k^{(N)})^{m-1}p_k^{(N)}&=&\cos\bigl((2k+1)\theta_N\bigr)^{2(m-1)}
\nonumber\\&&\times         \sin\bigl((2k+1)\theta_N\bigr)^2
\end{eqnarray}
for $m=1,2,3,\dots$~.

Each cycle queries the oracle $k$ times, once for each application of
$\mathcal{G}$, plus one more time during the test-state verification.
The verification is only done, however, if the result of the GA cycle is not
an index state to an oracle that is already known to be wrong from the
verification step of an earlier cycle.
If the search terminates after the $m$th cycle, the oracle has been queried as
many as
\begin{equation}
  \label{GAqueries}
  mk+1+(N-1){\left[1-{\left(\frac{N-2}{N-1}\right)}^{m-1}\right]}
\end{equation}
times on average, where the last summand is the average number of wrong test
states that are tried-out during the unsuccessful ${m-1}$ preceding cycles.

Accordingly, the average number that we need to query the oracle before we
know which oracle is the actual one, is given by
\begin{equation}
  \label{GQ-Nk}
G_Q(N;k)=\frac{k}{p_k^{(N)}}+\frac{N-p_k^{(N)}}{1+(N-2)p_k^{(N)}}\,.  
\end{equation}
This expression ignores the very small correction of no consequence 
that results from the possibility that the search can terminate after trying 
out ${N-1}$ test states for wrong oracles and so learning that the one 
remaining oracle must be the actual one. 
  
In $G_Q(N;k)$, 
$k$ is the number of oracle queries per cycle, so that we can optimize
GA by minimizing $G_Q(N;k)$ with respect to $k$,
\begin{equation}
  \label{GQ-opt}
  G_Q(N)=\min_kG_Q(N;k)\,.
\end{equation}
The asymptotic form of Eq.~(\ref{Gq}) is obtained from
\begin{equation}
  \label{GQ-lim}
  \lim_{N\to\infty}G_Q(N)/\sqrt{N}=\frac{\phi/2}{(\sin\phi)^2}=0.6900\,,
\end{equation}
where $\phi=1.1656$ is the smallest positive solution of ${2\phi=\tan\phi}$.
For ${N\gg1}$, one needs $(\sin\phi)^{-2}=1.18$ cycles on average before 
GA concludes successfully, and the optimal $k$ value is 
$k=\frac{1}{2}\phi\sqrt{N}=0.58\sqrt{N}$, which is slightly
less than 75\% of $k=\frac{1}{4}\pi\sqrt{N}$, the value that maximizes the
single-cycle success probability $p_k^{(N)}$.

\section{Alternative test-state search strategies}\label{sec:TSS-alt}
The GA search of Sec.~\ref{sec:GA-TSV} is consistently carried out in the full
space spanned by all index kets, as requested by the standard form of GA that
we accept as its definition.
By contrast, the successive iteration rounds of the test-state search of
Sec.~\ref{sec:TSA} are conducted in the relevant subspace spanned by the
remaining candidate index kets.  
As a consequence of this systematic shrinking of the searched space, the
successive educated guesses get better from one iteration round to the next.

In actual implementations, however, it may not be practical to limit the
search to the relevant subspace because it is usually much easier to realize
the necessary operations in the full $N=2^n$ dimensional space; 
see Sec.~\ref{sec:EXPT}.
If all iteration rounds of the test-state search are indeed performed in the
full space, we have
\begin{eqnarray}
  \label{GT-full}
  &&G_T'(N)=\frac{2-x}{1-x}-\frac{1}{N}\frac{1-x^N}{(1-x)^2}-\frac{1}{N}x^{N-2}
\nonumber\\&&\mbox{with}\enskip x=(N-1)\beta_{N-1}
\end{eqnarray}
instead of $G_T(N)$ of Eq.~(\ref{GT-recurr}).
The large-$N$ form thereof is
\begin{equation}
  \label{eq:GT-fullasym}
  G_T'(N)\approx\frac{e^{-\gamma}-1+\gamma}{\gamma^2}N=\frac{N}{6.08}
   \quad\mbox{with}\enskip\gamma=2+\sqrt{8}\,.
\end{equation}
Compared with the classical search, the reduction is still by more than a
factor of $3$, but the full-space test-state search needs about 12\% more
queries than the relevant-space search. 

One could wonder if there is a benefit in using the measurement for
unambiguous discrimination (MUD) \cite{MUD1,MUD2} rather than the SRM, 
because the MUD gives a 
small chance of identifying the actual oracle with a wrong test state. 
The probability of finding the right one of $N$ oracles with a randomly chosen
test state is then
\begin{equation}
  \label{TS-MUD1}
  \frac{1}{N}+\frac{N-1}{N}\frac{2}{N-2}=\frac{3N-4}{N(N-2)}
\end{equation}
where $2/(N-2)$ is the success probability for the MUD to the $(N-1)$-edged
pyramid of the $|t_j^k\rangle$ kets with ${j\neq k}$ \cite{pyramid3}.

The price for this increase of the bare $1/N$ probability is paid by getting
an inconclusive result from the MUD if it fails to identify the right state,
so that we have no information that would facilitate an educated guess for the
next test state.
The resulting average number of oracle queries is
\begin{equation}
  \label{TS-MUD2}
 G_T^{\mathrm{(MUD)}}(N)=\frac{(N-1)(3N+4)}{12N}
\end{equation}
if we successively search in the relevant subspace only, and
\begin{eqnarray}
  \label{TS-MUD3}
  &&{G_T^{\mathrm{(MUD)}}}'(N)=\frac{1}{1-x}-\frac{1}{N}\frac{x-x^{N+1}}{(1-x)^2}
                          -\frac{1}{N}x^{N-1}
\nonumber\\&&\mbox{with}\enskip x=\frac{N-4}{N-2}
\end{eqnarray}
if the search is consistently carried out in the full space.
The large-$N$ forms
\begin{eqnarray}
  \label{TS-MUD4}
  G_T^{\mathrm{(MUD)}}(N)&\approx&\frac{N}{4}\,,
\nonumber\\{G_T^{\mathrm{(MUD)}}}'(N)&\approx&\frac{N}{4/(1+e^{-2})}=\frac{N}{3.52}
\end{eqnarray}
show clearly that this price is high: The test-state search with MUD needs
substantially more oracle queries than the search with SRM.
In addition, the MUD is a proper POM and more difficult to implement than the
SRM.

One could also rely on the MUD rather than the SRM in the verification step 
of GA.
There are then modifications in Eqs.~(\ref{GAqueries}) and (\ref{GQ-Nk}), but
the large-$N$ statement of Eq.~(\ref{GQ-lim}) remains the same.

\section{Unitary operations for realizing the test-state approach}
\label{sec:EXPT}
While the test state $|t_{0}\rangle$ could be realized for any value of
$N$, we deal only with the important case of $N = 2^{n}$ when the oracles are
unitary operators acting on $n$ qubits.
Then, the test states $|t_{j}\rangle$ of Eq.~(\ref{tj}) are locally
equivalent to $|t_{0}\rangle$, in the sense that we
can transform the test state $|t_{0}\rangle$ into any other test state by
applying $X$ operations on the relevant qubits. 
In Sec.~\ref{sec:QCR} we describe a construction for $|t_{0}\rangle$, and show
how to realize the SRM of Eqs.~(\ref{DECOM})--(\ref{TjkTjl}) in
Sec.~\ref{sec:SRM}.

\subsection{Construction of the test state}\label{sec:QCR}
\begin{figure}
\centerline{%
\includegraphics[viewport=160 130 535 470,clip=true,width=0.9\columnwidth]%
{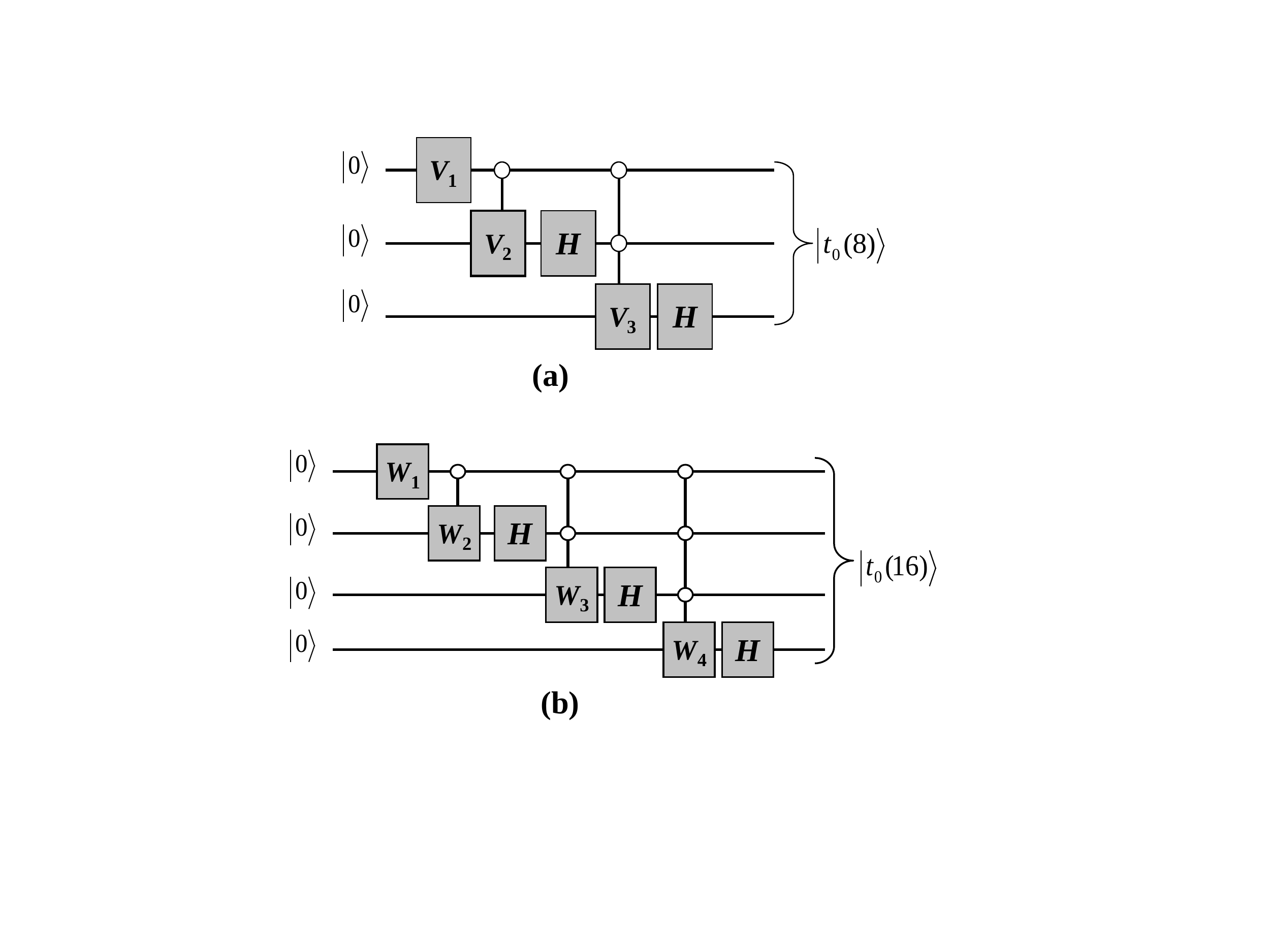}%
}
\caption{\label{fig:2}
  Quantum circuit (a) is for preparing of the three-qubit test state
  $|t_{0}(8)\rangle$ and (b) is for the four-qubit test state
  $|t_{0}(16)\rangle$, respectively. Here, the input state is
  $|0\rangle^{\otimes n}$ ($n=3, 4$), the Hadamard operations are depicted by
  $H$, and the explicit forms of the various controlled gates (the $V$s and
  $W$s) are given in the text, where all single-qubit operations are $Y$
  rotations.}  
\end{figure}

Let us first take the case of three qubits ($N = 8$) as an example; then 
$a=\sqrt{5/12}$, $b=\sqrt{1/12}$ in Eq.~(\ref{tj}) with $a^{2}+7b^{2}=1$. 
For preparing the three-qubit test state $|t_{0}(8)\rangle$, the input
register is initialized in the state $|0\rangle^{\otimes 3}$, and then the
single-qubit gate 
\begin{equation} \label{V1}
V_{1}=e^{-i\theta_1 Y}\quad\mbox{with}\enskip
\tan\theta_1=\frac{2b}{\sqrt{a^{2}+3 b^{2}}}=\frac{1}{\sqrt{2}}
\end{equation} 
is performed on the first qubit. 
Thereafter, we perform the controlled gate 
\begin{equation} \label{V2}
V_{2}=e^{-i\theta_2 Y}\quad\mbox{with}\enskip
\tan\theta_2=\frac{\sqrt{a^{2}+ b^{2}}- \sqrt{2}\,b}
                  {\sqrt{a^{2}+ b^{2}}+ \sqrt{2}\,b}=2-\sqrt{3}
\end{equation}
on the second qubit by taking the first qubit as control 
(with the control set to $|0\rangle$) 
followed by the Hadamard gate $H$ of Eq.~(\ref{H}). 
Subsequently, we perform the doubly-controlled gate 
\begin{equation}\label{V3}
V_{3}=e^{-i\theta_3 Y}\quad\mbox{with}\enskip
\tan\theta_3=\frac{a-b}{a+b}=\frac{3-\sqrt{5}}{2}
\end{equation}
on the third qubit by taking the first and second qubits as controls (with both
controls set to $|0\rangle$) followed by the Hadamard gate $H$.  
The over-all unitary operation $\textbf{u}$ for the case of three qubits can
be narrated as  
\begin{eqnarray} \label{u} 	
\textbf{u} & = & \!\bigl[I\otimes I\otimes H\bigr]\,
\bigl[|00\rangle\langle 00|\otimes V_{3}
+ \bigl(I\otimes I-|00\rangle\langle 00|\bigr)\otimes I\bigr]
\nonumber\\ &&\times
\bigl[I\otimes H\otimes I\bigr]\,
\bigl[\bigl(|0\rangle\langle 0|\otimes V_{2}
+ |1\rangle\langle 1|\otimes I\bigr)\otimes I\bigr]
\nonumber\\ &&\times
\bigl[V_{1}\otimes I\otimes I\bigr]\,,  	
\end{eqnarray}
and the corresponding quantum circuit is depicted in Fig.~\ref{fig:2}(a). 

The quantum circuit displayed in Fig.~\ref{fig:2}(b) is for the construction
of the four-qubit test state $|t_{0}(16)\rangle,$ where
$\textit{a}=\sqrt{13/28}$, $\textit{b}=\sqrt{1/28}$, and $a^{2}+15b^{2}=1$. 
In this case,
\begin{equation}\label{W1}
W_{1}=e^{-i\vartheta_1 Y}\quad\mbox{with}\enskip
\tan\vartheta_1=\frac{\sqrt{8}\, b}{\sqrt{a^{2}+7 b^{2}}}=\sqrt{\frac{2}{5}}
\end{equation} 
and
\begin{equation}\label{W2}
W_{2}=e^{-i\vartheta_2 Y}\quad\mbox{with}\enskip
\tan\vartheta_2=\frac{\sqrt{a^{2}+ 3b^{2}}- 2b}{\sqrt{a^{2}+ 3b^{2}}+ 2b}
=\frac{1}{3}
\end{equation}
as well as $W_{3}= V_{2}$ and $W_{4}= V_{3}$. 

The generalization to the $n$-qubit case is immediate. 
How to efficiently split a multi-qubit controlled unitary operation
$C^{n-1}(V)$ (single-qubit gate $V$ with ${n-1}$ control qubits) in terms of
universal gates with $n-1$ work qubits is shown in Refs.~\cite{gate, QCQI},
and its circuit complexity is of the order of $n$. 
Consequently, the circuit complexity for constructing the $n$-qubit test state
with quantum circuits of the kind shown in Fig.~\ref{fig:2} is $O(n^{2})$. 
In Appendix~\ref{sec:AM}, an alternative method for constructing the test
state $|t_{0}\rangle$ is given, where the amplitudes $a$ and $b$ are 
complex numbers.

\subsection{Realization of the SRM}\label{sec:SRM}
In order to perform the SRM of Sec.~\ref{sec:itrTSA} in the laboratory, 
one needs a unitary transformation
\begin{equation}
  \label{SRM-1}
  \textbf{M}=\sum_{l=0}^{N-1} |l\rangle\langle T_{0}^{l}|
\end{equation}
that turns each basis ket
$|T_{0}^{l}\rangle$ into the corresponding ket $|l \rangle$ of the
computational basis. 
With Eqs.~(\ref{tjj}) and (\ref{Tjk3}) we have 
\begin{eqnarray}
&&\textbf{M}=-\mathcal{I}+(1-a)|0\rangle\langle0| 
+b\bigl(|0\rangle\langle v|+ |v\rangle\langle0|\bigr)+
y|v\rangle\langle v|\nonumber\\
& &\mbox{with}\ |v\rangle=\sum_{k=1}^{N-1}|k\rangle\,, \label{M1} 
\end{eqnarray}
which has one eigenvalue $+1$ and ${N-1}$ eigenvalues $-1$, so that
the unitary operators $-\textbf{M}$ and the $n$-qubit-controlled-$Z$ 
\begin{equation}
C^{n-1}(Z)=\mathcal{I}-2|N-1\rangle\langle N-1|=\mathcal{O}^{N-1} \label{CZn}
\end{equation}
have the same set of eigenvalues, that is: they are unitarily equivalent.
The eigenkets of $\textbf{M}$ are
\begin{eqnarray}\label{Evector}
|e_{0}\rangle &=& \sqrt{\frac{1-a}{2}}|0\rangle 
+ \frac{b}{\sqrt{2(1-a)}}| v \rangle\,, \nonumber\\
|e_{1}\rangle &=& -\sqrt{\frac{1+a}{2}}|0\rangle 
+ \frac{b}{\sqrt{2(1+a)}}| v \rangle\,, \nonumber\\
|e_{j}\rangle &=& \frac{1}{\sqrt{2}}\bigl(-|1\rangle+ |j\rangle\bigr)
\quad\mbox{for}\enskip j= 2, 3,\dots,N-1\,,\qquad 
\end{eqnarray}
with $\textbf{M}|e_{0}\rangle = |e_{0}\rangle$ and 
$\textbf{M}|e_{j\neq0}\rangle = -|e_{j\neq0}\rangle$. 
In view of the degeneracy of $\textbf{M}$, the set of orthonormal eigenkets
for eigenvalue $-1$ is not unique, but the choice of Eqs.~(\ref{Evector}) is
particularly useful in the present context.
For, the eigenket $|e_{0}\rangle$ has the same structure as the test state
$|t_{0}\rangle$ of Eq.~(\ref{tj}), and we know from Sec.~\ref{sec:QCR} 
how to construct $|t_{0}\rangle$. 

\begin{figure}[t]
\centerline{%
\includegraphics[viewport=80 210 680 400,clip=true,width=0.9\columnwidth]%
{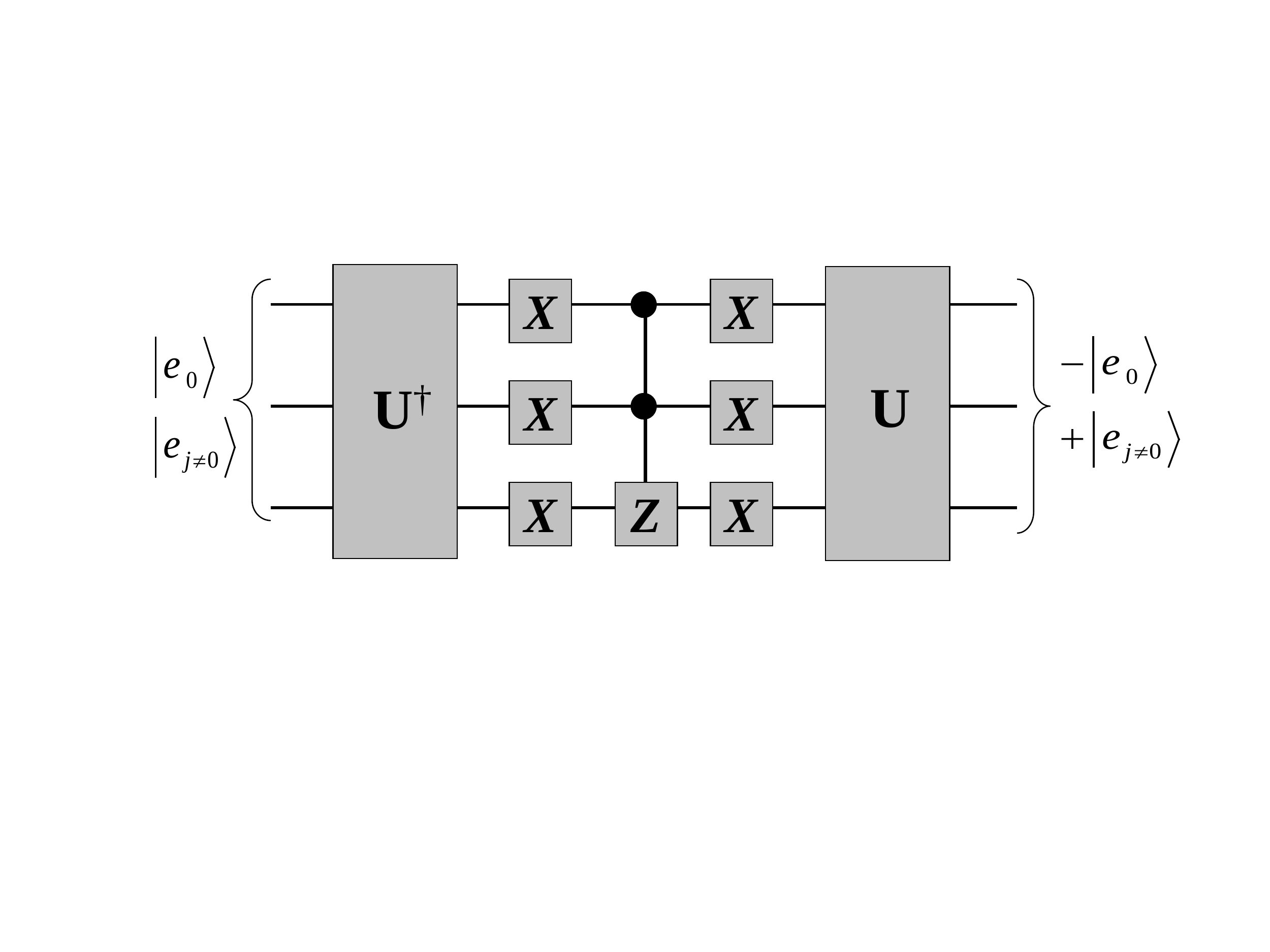}%
}
\caption{\label{fig:3}%
The quantum circuit for the implementation of $-\textbf{M}$ in the case of
three qubits. 
The operation $\textbf{U}$ is implemented by the circuit shown in
Fig.~\ref{fig:2}(a) after changing the parameters in accordance with
Eq.~(\ref{replace}). 
And, if we run the same circuit in the reverse order we can also implement
$\textbf{U}^{\dagger}$. 
The quantum gate shown in the center of circuit is the $C^{2}(Z)$ 
given by Eq.~(\ref{CZn}).}
\end{figure}

We relate $\textbf{M}$ to $C^{n-1}(Z)$ through the unitary operator
$\textbf{U}\,X^{\otimes n}$ that diagonalizes $\textbf{M}$ in the
computational basis,  
\begin{equation} \label{M}
\textbf{M}
=-\textbf{U}\,X^{\otimes n}\,C^{n-1}(Z)\,X^{\otimes n}\,\textbf{U}^{\dagger}
\end{equation} 
The operator $\textbf{U}$ itself is such that 
$\textbf{U}|0\rangle^{\otimes n}=|e_0\rangle$ or   
$\textbf{U}\,X^{\otimes n}|1\rangle^{\otimes n}=|e_0\rangle$, and we realize
it by the circuit for $\textbf{u}$---see Fig.~\ref{fig:2}---with the
replacements 
\begin{eqnarray} \label{replace}
a & \rightarrow & \sqrt{(1-a)/2},\nonumber\\
b & \rightarrow & b/\sqrt{2(1-a)}\,, 
\end{eqnarray}
while $\textbf{U}^{\dagger}$ is implemented by the circuit that has the gates
of Fig.~\ref{fig:2} in reverse order and all respective $\theta$ angle
parameters replaced by $-\theta$. 
Accordingly, all unitary factors on the right-hand side of Eq.~(\ref{M}) 
have known realizations, as illustrated for $N=2^3$ in Fig.~\ref{fig:3}.

With the SRM measurement thus implemented and the corresponding test states of
Sec.~\ref{sec:QCR}, we can verify the GA outcome and complete the quantum
search as discussed in Sec.~\ref{sec:GA-TSV}, and we can also perform the
full-space test-state search of Sec.~\ref{sec:TSS-alt}, for which
Eqs.~(\ref{GT-full}) and (\ref{eq:GT-fullasym}) apply.  
Of course, there are implementations as well of the test states in successively
smaller spaces and of the corresponding SRM measurements, but we are not aware
of economic implementations.
The restriction to the subspaces of yet-to-probe index states is 
rather awkward in practice.

\section{Summary and discussion}\label{sec:SummDiss}
We have introduced the test states that enable one to verify whether the
outcome of a quantum search with the Grover's algorithm is the actual oracle or
not.
We thereby regard the search problem as defined by the set of possible
oracles, which are those considered by Grover.
Other search problems, such as the one studied by H\o{}yer \cite{extGA}, are
not automatically covered as well; the corresponding test states---if they
exist---have to be found for each search problem separately.
That is also the case for Grover-type searches with more than one matching
item, that is when the oracle is a product of two or more different
unitary operators of the kind defined in Eq.~(\ref{Oj}). 

It is possible that there are no test states for some of these other search
problems, in which case one may not be able to verify if the search was 
successful --- neither by test states of some sort, nor by a method like the one
described in Appendix~\ref{sec:ACS} below, nor by another procedure.
This should make one wonder if a search problem is well-posed in the first
place, if it is not possible to verify the outcome. 
We leave this as a moot point.  

With the test states at hand, we have the option of solving the quantum search
problem with a classical search strategy.
But there is a twist: While there is one ``yes'' answer, each ``no'' answer is
slightly different and, with the help of the square-root measurement, 
this difference can be exploited systematically for a judicious 
choice of the test state for the next round.
This educated guessing is rewarded by much fewer queries of the oracle on
average than what one needs for the simple ``yes/no'' search.
A reduction by a factor of $3.41$ is achievable in principle, and a practical
scheme still gains a factor of more than three.
In our view, the classical-search benchmark is set by the search that exploits
the differences between the ``no''s fully. 

The picture is completed by giving explicit circuits for the implementation of
the $n$-qubit test states.
The circuit complexity is of order $n^2$, and a variant of the same circuit is
the main ingredient in the realization of the square-root measurement.

\begin{acknowledgments}
This work is supported by the National Research Foundation and the Ministry of
Education, Singapore. 
\end{acknowledgments}

\appendix

\section{An alternative confirmation step for Grover's algorithm}\label{sec:ACS} 
Here we describe an alternative procedure for verifying the result obtained by
GA. 
This method does not rely on the construction of test states.
Rather it employs a simple circuit 
that distinguishes between two selected ``target oracles'' 
and the other ${N-2}$ oracles. 
The verification is achieved by having the GA-outcome oracle in two different
target pairs, and thus requires two queries of the oracle.  

Suppose GA has had oracle $\mathcal{O}^j$ as the outcome.
The corresponding index ket $|j\rangle=|x_1\chi\rangle$ has value $x_1$ for
the first qubit and the values of qubits $2$,$3$, \dots, $n$ are summarized by
the string $\chi$. 
We pair $|j\rangle$ with $|\widehat{j}\rangle=|\widehat{x_1}\chi\rangle$ where
\begin{equation}
  \label{hat-x}
  \widehat{x_1}=x_1+1\ (\mbox{mod}\ 2)=\left\{
    \begin{array}{l@{\ \mbox{if}\ }l}
      1 & x_1=0\,,\\[1ex] 0 & x_1=1\,,
    \end{array}\right.
\end{equation}
so that $j$ and $\widehat{j}$ differ in the first bit value only.

\begin{figure}[t]%
\centerline{%
\includegraphics[viewport=170 210 580 380,clip=true,width=0.9\columnwidth]%
{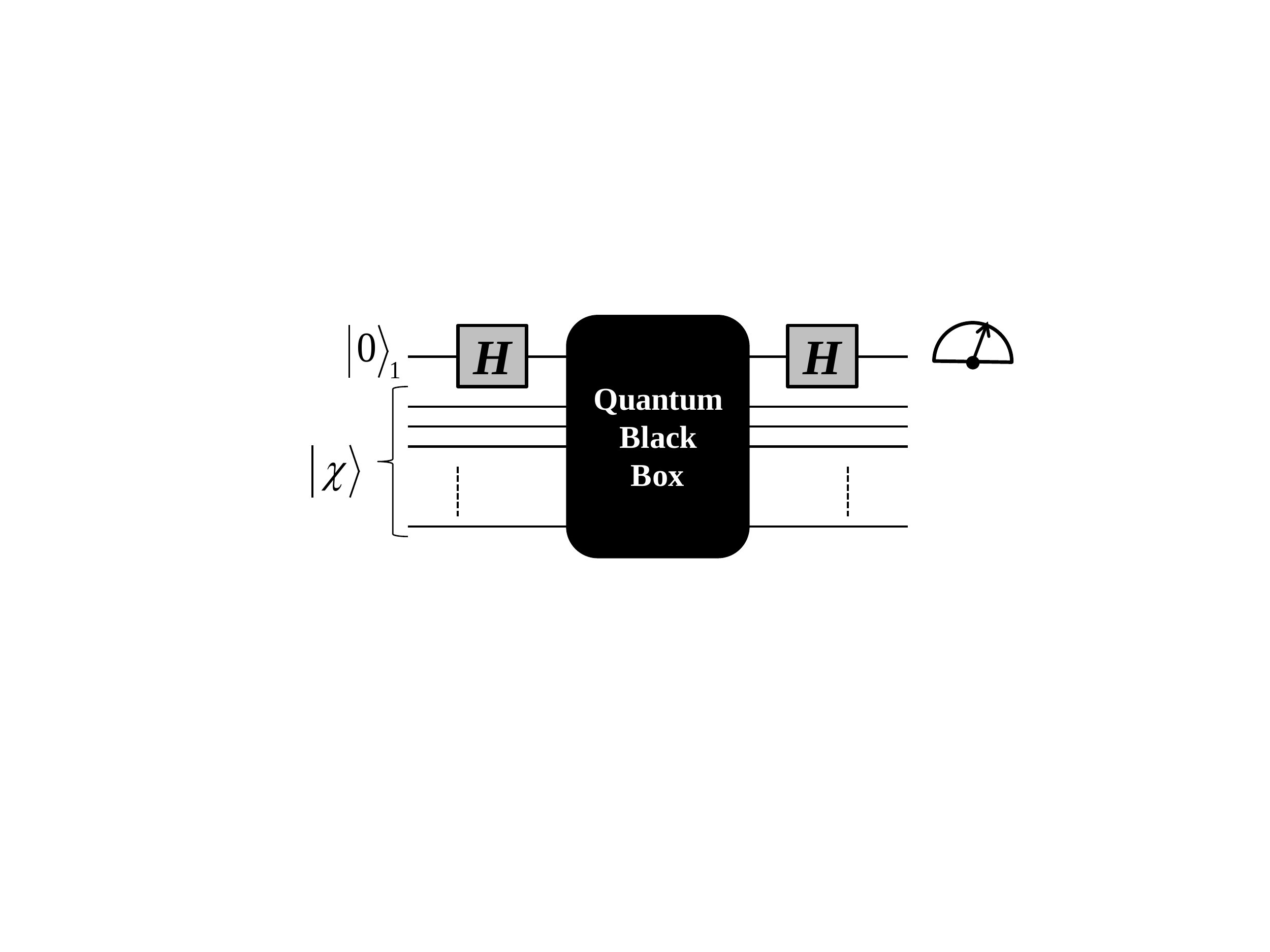}%
}
\caption{\label{fig:4}%
A single iteration of the alternative confirmation is exhibited in terms of
quantum circuit diagram. 
The input state with ket 
$|\phi_{\textrm{in}}\rangle=|0\chi\rangle$ of Eq.~(\ref{inState}) is passed
through the sequence of the Hadamard gate $H$ of Eq.~(\ref{H}), 
the quantum black box, and another Hadamard gate. 
Finally, the 1st qubit of the output state 
$|\phi_{\textrm{out}}\rangle$ is measured in the computational basis.}
\end{figure}

As indicated in Fig.~\ref{fig:4}, 
we prepare qubit~1 in the state with ket $|0\rangle$, and encode the 
$\chi$ part of the index state in qubits~$2$ through $n$.
So, the ket of the $n$-qubit input state is
\begin{equation}  \label{inState}
|\phi_{\textrm{in}}\rangle = |0\chi\rangle=\left\{
  \begin{array}{l@{\ \mbox{if}\ }l}
  |j\rangle & x_1=0\,, \\[1ex] |\widehat{j}\rangle & x_1=1\,.  
  \end{array}\right.
\end{equation}
We pass it through the quantum circuit of Fig.~\ref{fig:4}, 
where the given black box is used only once. 
If the black box is implementing either oracle $\mathcal{O}^j$ or oracle
$\mathcal{O}^{\widehat{j}}$, then the output state will have ket
\begin{equation} \label{outj} 
|\phi^{\mathrm{(yes)}}_{\textrm{out}}\rangle = |1\chi\rangle\,.
\end{equation}
If, however, the black box is implementing one of the other ${N-2}$ oracles, 
the output state will have ket  
\begin{equation}  \label{outk} 
|\phi^{\mathrm{(no)}}_{\textrm{out}}\rangle =|0\chi\rangle\,.
\end{equation}

Finally, qubit~1 is measured in the computational basis. 
If we find $0$, the ``no'' output is the case, and we can be
sure that the actual oracle is neither $\mathcal{O}^j$ nor
$\mathcal{O}^{\widehat{j}}$.
But when we find $1$, we know that one of these oracles is inside
the black box.
We determine which one by pairing $|j\rangle$ with a third index ket that also
differs only by the value of one qubit, which then plays the role of the
privileged qubit in the corresponding circuit of the kind depicted in 
Fig.~\ref{fig:4}, where qubit~1 is singled out. 

So, we either get a definite ``no'' answer to the question ``Is the
$j$th or the $\widehat{j}$th oracle the case?'' or we are told ``yes, it is
one of these two.'' 
In the latter situation, we know for sure which one it is after a second round.

\section{An alternative construction of the test states}\label{sec:AM} 
In Sec.~\ref{sec:QCR}, we gave a construction of the test states of
Eq.~(\ref{tj}) with the real coefficients $a$ and $b$ of Eq.~(\ref{ab}).
Here, we provide an alternative method by which one produces the alternative
test states with complex $a$ and $b$ amplitudes,
as exemplified by 
\begin{eqnarray}  \label{t0}
|t_0\rangle &=& a|0\rangle+b\sum_{l=1}^{N-1}|l\rangle\nonumber\\
&=& (a-b)|0\rangle^{\otimes n}+b\sqrt{N}|+\rangle^{\otimes n}\,,
\end{eqnarray}
where $|+\rangle=H|0\rangle=\bigl(|0\rangle+|1\rangle\bigr)/\sqrt{2}$ uses the
Hadamard gate of Eq.~(\ref{H}), and the
absolute values $|a|$ and $|b|$ are, of course, still those of Eq.~(\ref{ab}).
As before, it is enough to show how $|t_0\rangle$ is made, the other test
states are then available by applying some single-qubit $X$ gates.

We obtain a ket of this kind by applying the multi-Hadamard unitary operator 
\begin{equation} \label{nH}
U^{12\dots n}_{HH...H}(\theta) = \exp\bigl(-i\theta H^{\otimes n}/2\bigr)\,,
\end{equation}
to  $|j=0\rangle=|0\rangle^{\otimes n}$,
\begin{equation}\label{psi} 
 U^{12\dots n}_{HH\dots H}(\theta)|0\rangle^{\otimes n}
= \cos\frac{\theta}{2}\,|0\rangle^{\otimes n}
   -i\sin\frac{\theta}{2}\,|+\rangle^{\otimes n} \,.
\end{equation}
Now, for $b\sqrt{N}=-i\sin(\theta/2)=-i\sqrt{N/(2N-4)}$ we need to set the angle
parameter $\theta$ to the value determined by
\begin{equation}  \label{theta}
\tan\frac{\theta}{2} = \sqrt{\frac{N}{N-4}}\,,
\end{equation}
and one verifies that 
\begin{equation}
  a= \cos\frac{\theta}{2}-\frac{i}{\sqrt{N}}\sin\frac{\theta}{2}
   =\frac{\sqrt{N-4}-i}{\sqrt{2N-4}}
\end{equation}
also has the absolute value required by Eq.~(\ref{ab}).
So, if we set $\theta$ in accordance with Eq.~(\ref{theta}), then the output
state of Eq.~(\ref{psi}) is the test state $|t_0\rangle$ of Eq.~(\ref{t0}). 
We note that $\theta=\pi$ for $N=4$, and $\theta=\pi/2+2/N$ for $N\gg1$.

\begin{figure}[t]
\centerline{%
\includegraphics[viewport=60 150 665 410,clip=true,width=0.9\columnwidth]%
{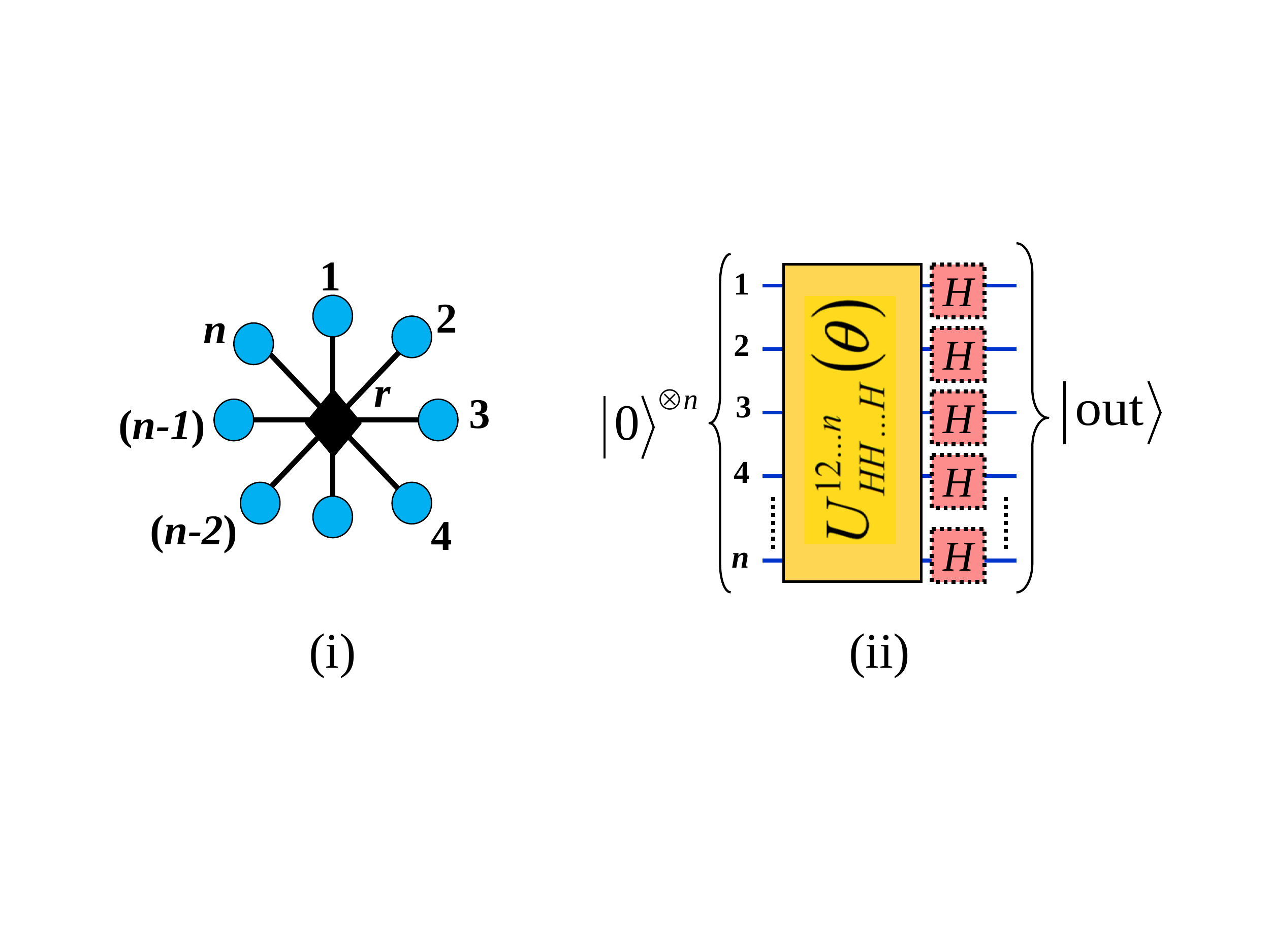}%
}
\caption{\label{fig:5}%
(Color online) (i) The ``star graph'' stands for the graph state with the ket
$|\phi\rangle_{(1+n)}$ of Eq.~(\ref{gr2}). 
The (blue) circles represent the $n$ qubits which carry the input ket 
$|0\rangle^{\otimes n}$; the bonds are established by the controlled-Hadamard
gates $\textsc{ch}(n)$ of Eq.~(\ref{chn}), 
and the ancilla qubit `$r$' is represented by the (black) diamond. 
Plot~(ii) represents the net effect on the input ket $|0\rangle^{\otimes n}$, 
when the ancilla qubit is measured in an appropriately chosen basis.}
\end{figure}

One can execute the unitary operation $U^{12\dots n}_{HH\dots H}(\theta)$ on
the $n$-qubit input state $|0\rangle^{\otimes n}$ by a similar method as the
one given for the unitary operation $U^{12\dots n}_{ZZ\dots Z}(\theta)$ 
in  Sec.~IIA of Ref.~\cite{HQCM}.  
Here, the input quantum register of $n$ qubits (circles in
Fig.~\ref{fig:5}(i)) and the ancilla qubit `$r$' (diamond in
Fig.~\ref{fig:5}(i)) are initialized in the $n$-qubit input state with ket 
$|0\rangle^{\otimes n}$ and the state with ket $|+\rangle_r$, respectively. 
Then, similar to the $n$ $\textsc{cz}$ operations in Sec.~II\,A
of Ref.~\cite{HQCM}, here we perform the $n$ controlled-Hadamard operations
\begin{equation}\label{chn}
\textsc{ch}(n)= \bigl(|0\rangle\langle 0|\bigr)_r\otimes I^{\otimes n}
+\bigl(|1\rangle\langle 1|\bigr)_r\otimes H^{\otimes n}
\end{equation}
between the ancilla qubit and each one of the $n$ qubits. 
All the controlled-Hadamard operations represented by the bonds in  
Fig.~\ref{fig:5}(i) can be carried out at the same time, because they all
commute with each other. 
This leads us to the resultant star-graph state with the ket  
\begin{equation}
|\phi\rangle_{(1+n)} = \frac{1}{\sqrt{2}} 
\bigl(|0\rangle_{r}\otimes|0\rangle^{\otimes n} 
+ |1\rangle_{r}\otimes|+\rangle^{\otimes n}\bigr)\,.
\label{gr2}
\end{equation}
The subscript $1+n$ reveals the number of qubits of the final graph state.

A single-qubit projective measurement on the ancilla qubit `$r$' in the basis 
\begin{eqnarray} \label{bas}
|\!\uparrow(\theta,\tfrac{1}{2}\pi)\rangle_{r} 
& = & \cos\frac{\theta}{2}\,|0\rangle_{r} 
+ i\sin\frac{\theta}{2}\,|1\rangle_{r}\,,   \nonumber \\
|\!\downarrow(\theta,\tfrac{1}{2}\pi)\rangle_{r} 
& = & - \sin\frac{\theta}{2}\,|0\rangle_{r} 
+ i\cos\frac{\theta}{2}\,|1\rangle_{r}
\end{eqnarray}
transforms the input ket of the $n$ qubits into the ket  
\begin{equation}
|\mathrm{out}\rangle=\bigl(H^{\otimes n}\bigr)^{m_{r}}
U^{12\dots n}_{HH\dots H}(\theta)|0\rangle^{\otimes n}\,.
\label{out2}
\end{equation}
Here, $m_{r}\in\left\{0, 1\right\}$ is the measurement result, 
and $\bigl(H^{\otimes n}\bigr)^{m_{r}}$ is the byproduct operator 
\cite{Raussendorf01, Raussendorf011}, which is represented by the dotted boxes
on all the $n$ qubits in Fig.~\ref{fig:5}(ii). 

After undoing the effect of the byproduct operator in Eq.~(\ref{out2}), one has
the test state of Eq.~(\ref{t0}), and can then apply the necessary
single-qubit $X$ gates to get the test state that one needs.
Alternatively and more efficiently, one can combine these $X$ gates with the
byproduct operator and execute the resulting single-qubit gates in one go.

\end{document}